\documentclass[11pt]{article}
\usepackage{pifont}
\usepackage{natbib}
\usepackage{graphicx}
\usepackage{amssymb}
\usepackage{epstopdf}
\usepackage{amsmath}
\begin{document}

\begin{center}
{\bf {\large {Comparative simulations of the 2011 Tohoku tsunami with MOST and Cliffs}}}\\\vspace{1.6mm}
Elena Tolkova$^1$\\\vspace{1mm}
{\emph{$^1$NorthWest Research Associates}}
\end{center}
\bigskip

{\bf{Background}}\\

The Method of Splitting Tsunamis (MOST) is a depth-averaged long wave tsunami inundation model adapted by the National Oceanic and Atmospheric Administration (NOAA) for tsunami forecasting operations \cite[]{titov2005}.
Presently, the MOST model is incorporated into the Short-term Inundation Forecast of Tsunamis (SIFT) system, which also integrates real-time deep-ocean observations of tsunamis, a basin-wide pre-computed propagation database of unit tsunami wave-fields representing hypothetical unit earthquakes, and sets of high-resolution grids focusing on specific coastal locations.
Once an open-ocean estimate of the tsunami wave in terms of the database unit tsunamis is obtained, live modeling with MOST is used to provide tsunami inundation forecasts for the coastal locations, under the boundary input from the propagation database \cite[]{tang2012}.

\emph{As can be noticed from the comparisons of modeled and recorded time-histories of recent tsunamis (see Event Pages at 
http://nctr.pmel.noaa.gov/ database\_devel.html),
the forecast often underestimates later waves.} \\

Official home to the MOST model is the NOAA's Pacific Marine Environmental Laboratory (PMEL), where few versions of the model exist, with the last version known as MOST-4. Wang et al. (2013) demonstrated that MOST-4 "improves ... the model accuracy for later waves" compared to the earlier versions. Specifically, MOST-4 "enhances the later wave simulation, through the improvement to the reflective boundary conditions in the code" \cite[]{wang_agu}.\\

The later improvement, however, has not addressed the entirety of the problem.  A different solution on a wet-dry boundary is highlighted here with a few comparative simulations. The new solution, termed Cliffs, exceeds all MOST versions in accuracy of computing later  waves. This is demonstrated with simulation of the Tohoku-2011 tsunami to Monterey Bay, CA, and into fiords, bays, and inlets of southeastern Alaska, followed by comparison with tide gage records.\\

{\bf{Simulation of Tohoku-2011 tsunami from the source earthquake to Monterey Bay}}\\

The March 11, 2011 Tohoku tsunami was simulated in the Pacific from the source earthquake with MOST-4 and Cliffs. The tsunami source is the one used in PMEL \cite[]{tang2012}. As adapted for ocean-wide computations in PMEL, simulations were carried at 4 arc-min resolution in longitude and 
latitude, with a reflecting wall at 20 m deep. 
Next, the ocean-wide solution was refined with two nested grids narrowing on Monterey bay, California. The outer grid covered the Northern California coast, while the finer inner grid at 8 arc-sec enclosed the bay and its surroundings.  
\begin{figure}[ht]
	\resizebox{\textwidth}{!}
		{\includegraphics{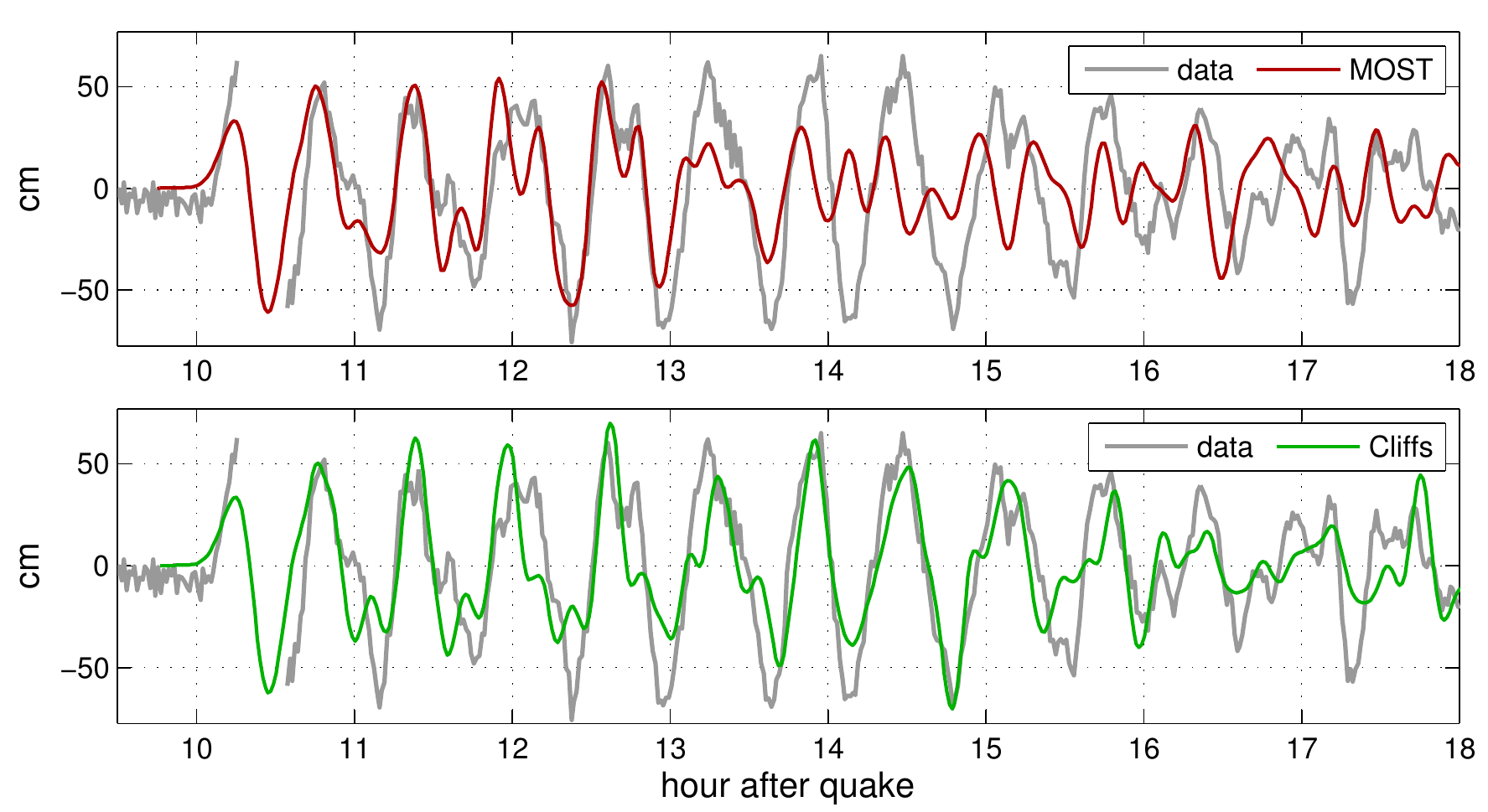}} 
	\caption{Top: Tohoku-2011 tsunami at Monterey Bay tide gage, recorded (gray) and modeled from the source with MOST-4 (red); bottom: the same, modeled with Cliffs (green). All simulations are delayed 8 min.}
	\label{tohoku2}
\end{figure}
\begin{figure}[ht]
	\resizebox{\textwidth}{!}
		{\includegraphics{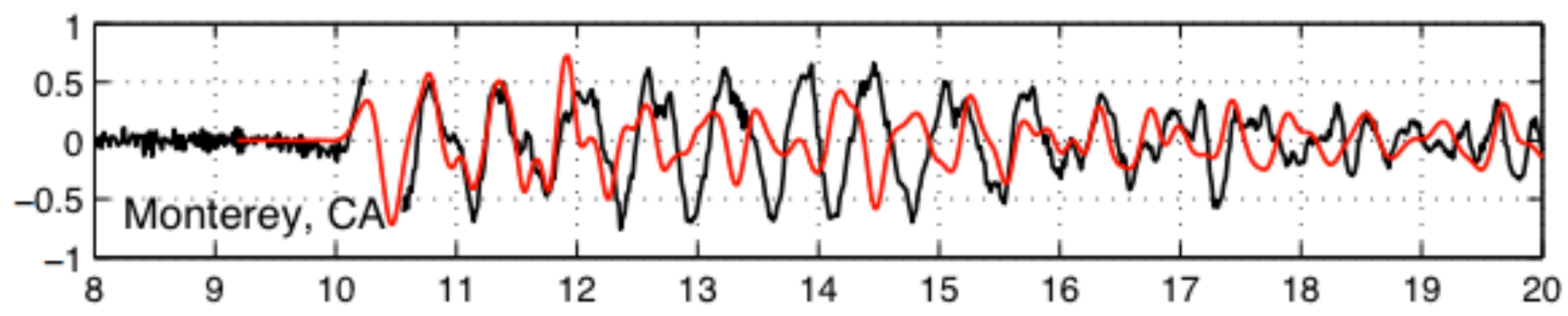}} 
	\caption{Tohoku-2011 tsunami at Monterey Bay tide gage, recorded (black) and modeled in PMEL with SIFT (red), delayed 3 to 15 min.  Courtesy of NOAA/PMEL/Center for Tsunami Research.}
	\label{pmel}
\end{figure}

Figure \ref{tohoku2} shows the Tohoku-2011 tsunami time histories, recorded at the tide gage and simulated with either MOST-4 or Cliffs. A simulation performed with MOST-4 matched the record's envelope well (subject to limited knowledge about the tsunami source function) for about 3 hrs, until a 13 h mark. A simulation performed with Cliffs matched the record for twice as long, until a 16 h mark. 
The better results by Cliffs is due to the better representation of later waves in the ocean-wide simulation.
Very little difference in the results by MOST-4 and Cliffs in Monterey Bay is observed, should the regional simulations use the same boundary input from the Pacific-wide simulation. 

For comparison with the earlier MOST versions, Figure \ref{pmel} shows the tsunami at the Monterey gage simulated in the PMEL with the SIFT system in 2011, reproduced from http://nctr.pmel.noaa.gov/honshu20110311/ images/comp\_plots /2011honshu \_ westcoast.pdf. This simulation under boundary input from the PMEL propagation database was able to match the record for no longer than 2 h, until a 12 h mark.\\

{\bf{Simulation of Tohoku-2011 tsunami in southeast Alaska}}\\

\begin{figure}[ht]
\begin{tabular}{c}
	\includegraphics[height=0.6\textwidth]{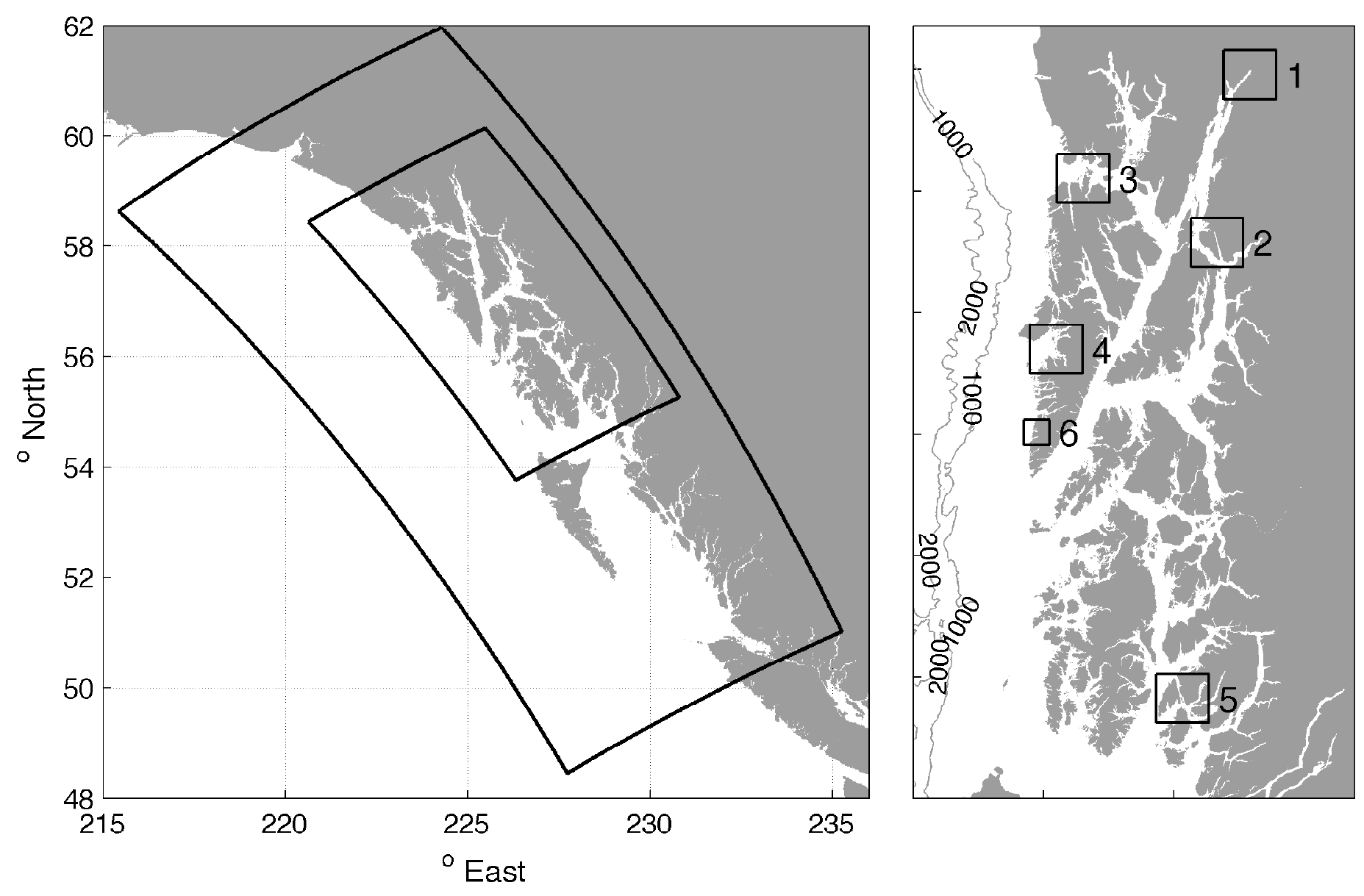} \\
	\includegraphics[height=0.6\textwidth]{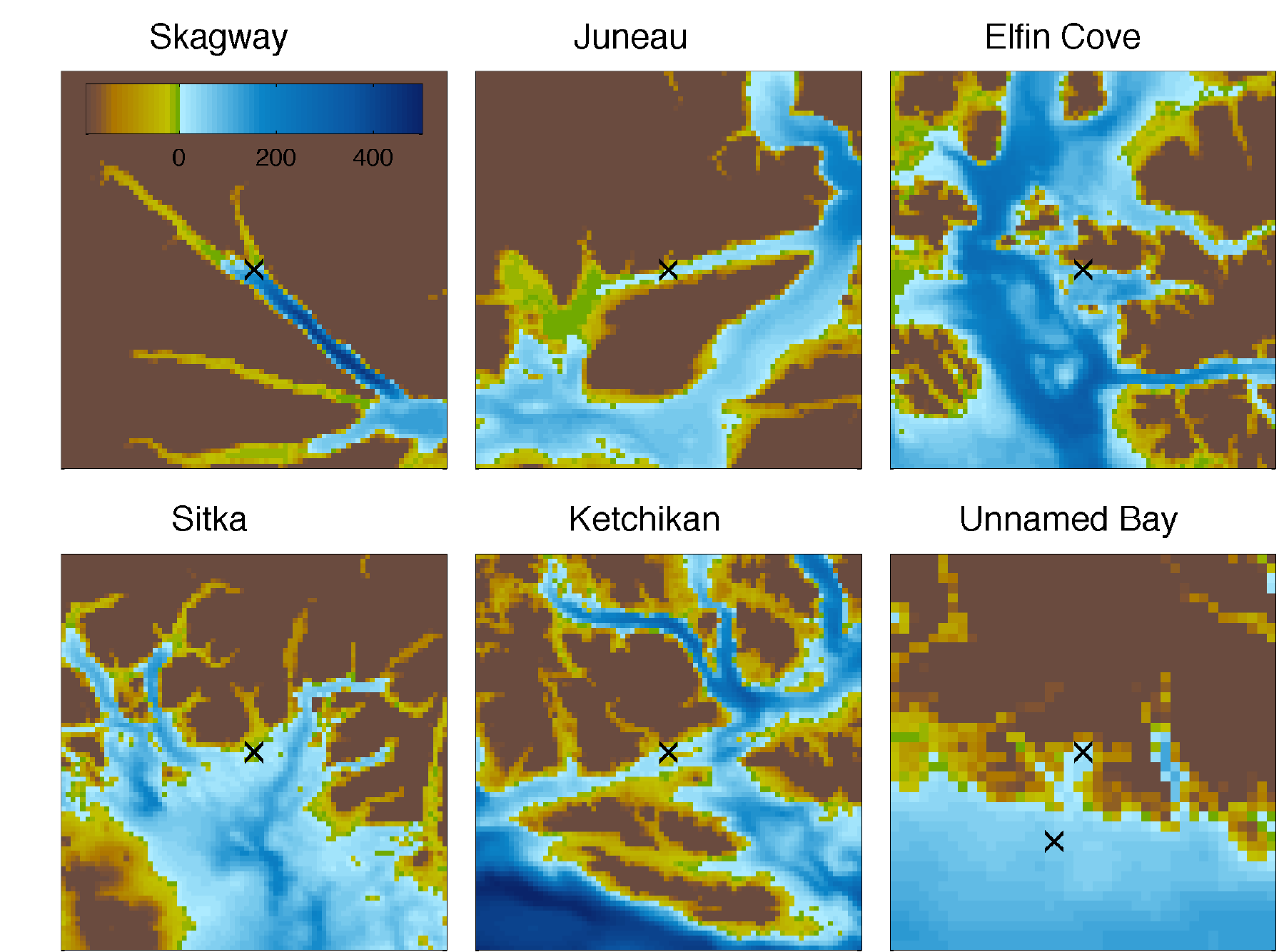}
\end{tabular}
	\caption{Top left: extent of two nested regional level grids for the Tohoku tsunami simulations in southeast Alaska. Top right: the inner regional grid; boxes denote areas zoomed-in under. Bottom: $40 \times 40$ km areas around tide gages (crosses) in  1- Skagway, 2 - Juneau, 3 - Elfin Cove, 4 - Sitka, 5 - Ketchikan at 16 arc-sec (0.5 km) space resolution.}
	\label{ak1}
\end{figure}

\begin{figure}[ht]
	\resizebox{1.1\textwidth}{!} %
		{\includegraphics{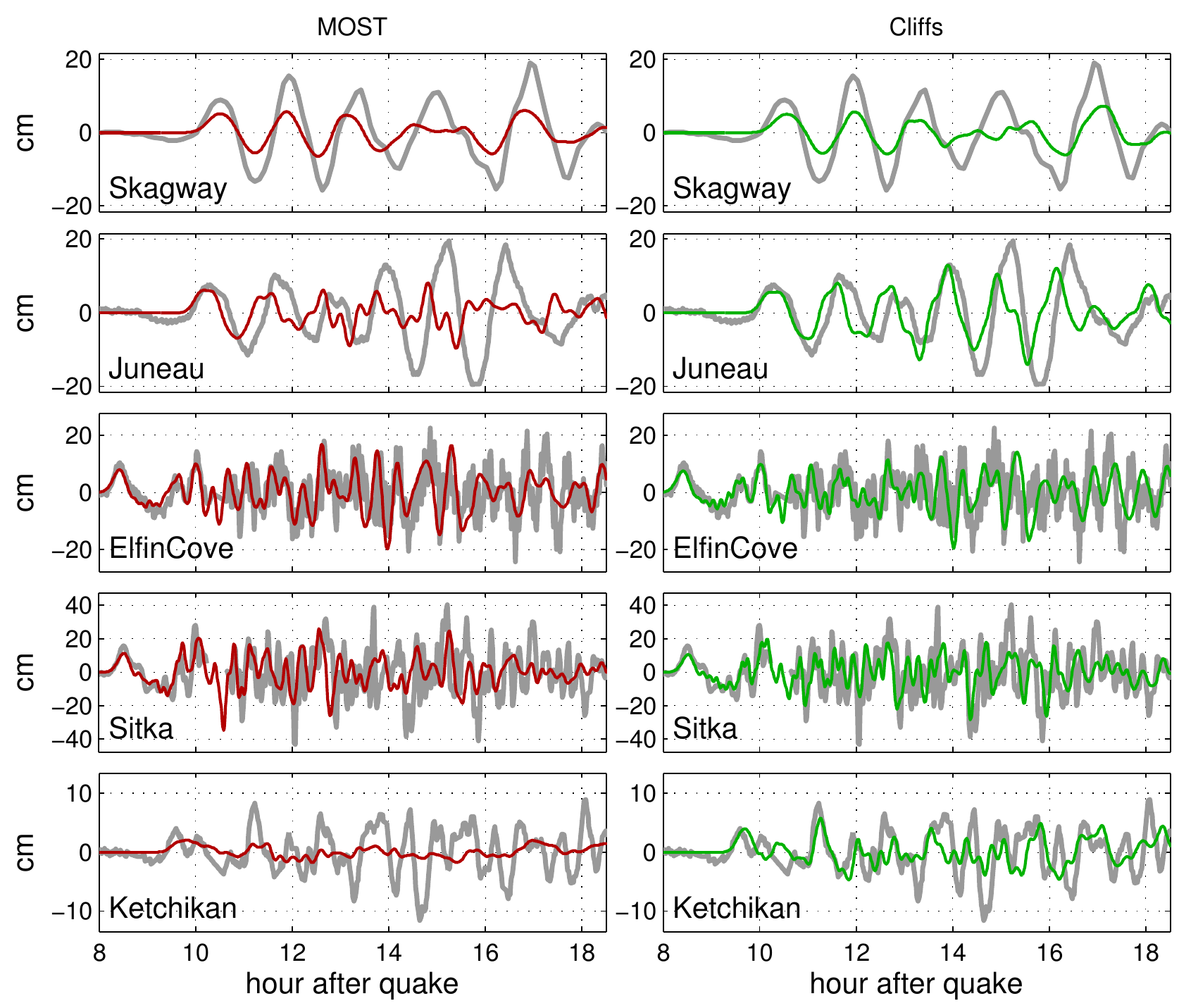}} 
	\caption{Simulated (color) and recorded (gray thick), tide removed, tsunami time-histories at NOS tide gages; left - computed with MOST-4, right - computed with Cliffs. All simulated time histories are delayed 8 min.}
	\label{ak3}
\end{figure}

The difference in regional simulations by MOST-4 and Cliffs might appear even under the same boundary input from the Pacific. Below, the 2011 Tohoku tsunami is simulated with MOST-4 and Cliffs in southeast Alaska. The regional grids extent is shown in Figure \ref{ak1}. Computations at the regional level were performed with MOST-4 and with Cliffs, {\large{\emph{under the same boundary input from the Pacific-wide simulation of the tsunami  computed with Cliffs}}}. 
The simulation results are compared location-wise between the two models and with local tide gage records. 

The Tohoku tsunami has been recorded at six tsunami-capable NOS gages throughout the area (Figure \ref{ak1}). 
According to gage records (Figure \ref{ak3}), wave motion at Skagway, Juneau, and Ketchikan is dominated by components with periods longer than 20 min, which in depth 50 m corresponds to wavelengths longer than 28 km. There is some presence of shorter-period waves at Sitka and Elfin Cove, which face the ocean more directly. Given the inner regional grid resolution 0.5 km, we expect to capture the major part of tsunami signal at all locations except Port Alexander where the record is heavily dominated by an oscillation with 9 min period. 
Since the recorded tsunami amplitude was well under 1 m at any gage, little of runup/rundown action could take place which, together with a relatively coarse resolution, justifies for using reflective boundary conditions on the shoreline. 

Simulated time histories at the gages computed with MOST-4 (left) and Cliffs (right), and the de-tided gage records are shown in Figure \ref{ak3}. The solutions by MOST-4 and Cliffs display fairly minor differences in Skagway, Sitka, and Elfin Cove. However, there are significant differences between the models at Ketchikan and Juneau. Cliffs solution matched the gage records as closely as in other locations, subject to the accuracy of the tsunami source estimate. MOST-4 clearly was not able to propagate the wave through the narrow channels without loss of the signal. 

\end{document}